# Design of Macroscale Optical Systems with Metaoptics Using Transformer-Based Neural Networks

Ryan C. Ng[1], Stéphane Larouche[1], Peter Y. Schneider[1], Aditi Munshi[1], Robert Bedford[2], Philip W. Hon[1], Katherine T. Fountaine[1]

1. Northrop Grumman Systems Corporation, Research Laboratories, 1 Space Park Drive, Redondo Beach, CA. 90278
2. Materials and Manufacturing Directorate, Air Force Research Laboratory, Wright-Patterson Air Force Base, Dayton, Ohio 45433, USA

**Abstract**

Metaoptics are thin, planar surfaces consisting of many subwavelength optical resonators that can be designed to simultaneously control the amplitude, phase, and polarization to arbitrarily shape an optical wavefront much in the same manner as a traditional lens but with a much smaller form factor. The incorporation of metaoptics into a conventional optical system spans multiple length scales between that of the individual metaoptic elements ($< \lambda$) and that of the entire size of the optic ($>> \lambda$), making computational techniques that accurately simulate the optical response of metaoptics computationally intractable, while more efficient techniques utilizing various approximations suffer from inaccuracies in their prediction of the optical response. To overcome the trade between speed and accuracy, we implement a transformer-based neural network solver to calculate the optical response of metaoptics and combine it with commercial ray optics software incorporating Fourier propagation methods to simulate an entire optical system. We demonstrate that this neural net method is more than 3 orders of magnitude faster than a traditional finite-difference time domain method, with only a 0.47 % deviation in total irradiance when compared with a full wave simulation, which is nearly 2 orders of magnitude more accurate than standard approximation methods for metaoptics. The ability to accurately and efficiently predict the optical response of a metaoptic could enable their optimization, further accelerating and facilitating their application.

**Main**

In a standard optical element, the material's refractive index and the shape of its surface(s) control the scattering of light. Unfortunately, as the clear aperture of the optical element grows, so does the mass and volume needed to maintain optical and mechanical requirements. Metaoptics, by contrast, are planar surfaces with judiciously placed subwavelength nanostructured optical resonator elements that enable arbitrary wavefronts through independent and simultaneous control of the phase, amplitude, and polarization of light via tuning of the nanostructure geometry rather than the shape of the entire lens. [1-4] Metaoptics can shape an optical wavefront in a wafer-thin and lightweight form factor to reduce cost, size, and weight for optical systems. There are a significant number of relevant and interesting applications for metaoptics, such as in imaging, sensing, communications and beamsteering. [5-7] Despite their relevance and potential, design and accurate simulation of metaoptics, typically done with finite-difference time domain (FDTD) or finite element method solvers, is complex and computationally expensive due to their nanoscale features and macroscale extent, which has challenged their development and incorporation.

Currently, optical systems are commonly designed in ray optics software such as Zemax OpticStudio, with built-in Fourier optics propagation methods (angular spectrum or Fresnel) for beam propagation through free space or an optical chain of elements. These tools are not inherently designed to model metaoptic elements, but as metaoptic research has progressed in its technology readiness level, there is an increasing need for an accurate metaoptic design tool that integrates well with existing ray optics software. To date, researchers have generally relied on three methods for modeling a metaoptic in this situation (see **Figure 1**), with some trade-off between speed and accuracy, often sacrificing accuracy for a computationally tractable metaoptic simulation. With each of these methods, an incoming incident electric field propagated via Fourier optics propagation is the input for said method to predict the field leaving the metaoptic, which is then propagated onwards, again, with Fourier optics propagation. Within OpticStudio's geometric optics (GO) environment, a metaoptic can be modeled assuming idealized meta-atom performance with continually varying, full $2\pi$ phase control and unity transmission, which is not realistic and has limited accuracy. A second improved option that is commonly implemented, also known as the local phase approximation (LPA), [8-10] uses a meta-atom library with unit cell scattering responses (phase, transmission) based on full-wave simulations of infinite arrays; this method improves accuracy over the GO OpticStudio method because it accounts for the varying, non-unity transmission of each meta-atom while also capturing the discrete nature of the metaoptic phase profile, but it still exhibits accuracy issues for larger and faster optics because it does not account for meta-atom nearest neighbor interactions, which is particularly critical at Fresnel zone edges. Non-idealities in the transmission cause undesired diffraction orders and must be considered to correctly predict the performance of the metaoptic. As of 2024, OpticStudio also includes the ability to account for non-ideal transmission scattering within their GO environment, but it does not account for neighbor interactions. [8] The third method is a full-wave simulation, using numerical solvers such as FDTD, rigorous coupled-wave analysis (RCWA) or finite element method (FEM). This is the most accurate method but is computationally expensive due to the orders of magnitude difference between the length scale of the metaoptic elements and the entire optic. In general, full-wave simulations of full-size optics (~10,000+ $\lambda$ in diameter) is not tractable both in terms of required memory and compute time. The aforementioned three methods are assessable through Ansys' recent offerings of integrated tools (RCWA, FDTD, and OpticStudio). [8,11]

Here, we propose an alternative design method that accurately simulates the optical performance of a system incorporating metaoptics, while being orders of magnitude faster than traditional FDTD (weeks or months reduced down to minutes) with accuracy approaching that of a full-wave simulation of the entire lens. This process flow is illustrated in **Figure 1**, along with the three previously mentioned methods. The general approach for integration of metaoptics into traditional optical modeling systems consists of propagation through the optical chain by a field propagation method, e.g. OpticStudio's physical optics propagation (POP) tool, which uses Fourier propagation methods, and a hand-off to a separate calculation method for the metaoptic. For the hand-off, upon arriving at the metaoptic, the electric field is exported from OpticStudio's POP tool, and then passed through a metaoptic calculation method (LPA, neural network, full-wave), which is subsequently imported back into OpticStudio's POP tool for continued propagation through the optical chain. This same procedure can be extended to any number of metaoptics within the optical chain, and calculation of performance of the entire chain can easily be handled by interfacing with OpticStudio via its Python API.

For this new method, we use a transformer-based neural network (NN) [12,13] that serves as a surrogate electromagnetic solver capable of accurately and efficiently predict the performance of metaoptics. The trained NN is specific to our selected meta-atom library (cylindrical air holes in silicon) and wavelength (4 μm), but the method can be readily generalized to any material/geometry/wavelength system. Additional details on the meta-atom library are included in the Supporting Information. To train the NN, we simulated many small converging and diverging metaoptics in Lumerical FDTD with f-numbers ranging from 1 to 10, as well as some metaoptics with sinusoidal phase profiles to capture the full range of radial trends possible for a general optic. The NN was trained to predict the field directly after each individual meta-atom considering the applied field, the meta-atom, and its neighboring meta-atoms. The field is then stitched together to capture the behavior of the full metaoptic; this structure enables scaling to large optics with minimal computational consequence. We explored multiple NN types/configurations and neighborhood sizes; optimal performance was found for a 15x15 neighborhood of meta-atoms built from a 5x5 array of 3x3 nanopillar arrays, with each 3x3 array serving as a token for the transformer-based model (see Supporting Information for more details on training). Consequently, the resultant NN predicts output electric fields while considering effects of neighboring pillars, which implies that in addition to typical nearest neighbor effects (slowly varying meta-atom radii), the NN method also accurately captures behavior at the Fresnel zone edges (abrupt changes in the meta-atom geometry).

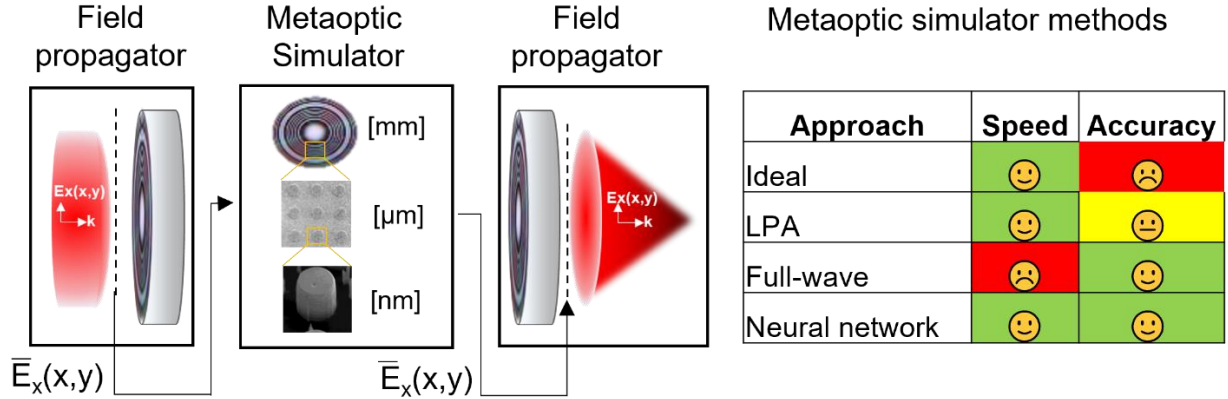

**Figure 1.** Process flow for metaoptic design methods. Field propagation is handled via OpticStudio's physical optics propagation tool with complex fields before the metaoptic as input into various metaoptic model methods, each with various trade-offs between speed and accuracy. The transformer-based NN surrogate electromagnetic solver integrated with OpticStudio's

physical optics propagation tool that we propose in this work has a combination of short simulation time as well as accuracy.

Following training, the NN quickly and accurately predicts the electric field after each individual nanopillar while accounting for the effect of neighboring pillars. A previous paper from Zhelyeznyakov et al. used a similar method with a NN for metaoptic performance prediction. [14] Their method differs in a few ways – (1) they used a single value decomposition to reduce the fields to a single dimension followed by a deep NN, (2) they used only immediate nearest neighbors (3x3 meta-atoms centered on the atom considered), (3) they discarded the relative positional information of the nanopillars by flattening the inputs and feeding them into a dense layer, and (4) they did not include a method for incorporating the NN prediction tool into an optical system, i.e. the NN only predicted single metaoptic performance with normal incidence illumination. We believe that our transformer-based method with complex field prediction, use of 15x15 meta-atoms, and integration with larger optical systems (described later in the text), represents a significant advancement in the use of machine learning methods for metaoptic performance prediction.

**Figure 2** compares the prediction accuracy of the NN method relative to ideal, LPA, and full-wave calculation. The ideal method assumes unity amplitude for the meta-atoms, while LPA uses the meta-atom transmission based on simulation of individual meta-atoms in an infinite periodic lattice. In **Figure 2**a and b, we compare the field (phase and field intensity) directly after the metaoptic and at the focal point cross section, respectively, for an f/2.5 metaoptic with a diameter of 250 μm. The f/2.5 metaoptic was selected for this comparison because it was not included in the NN training data and is therefore an appropriate structure for validation. From these comparisons, we note that the ideal method has particularly poor prediction power for metaoptic behavior, as expected; the LPA prediction is able to provide an estimation for the metaoptic optical performance but falls short of fully capturing field behavior and subsequently misrepresents the ultimate power at the focal plane. In particular, the LPA method struggles to accurately capture behavior at Fresnel zone edges (where phase wraps from -$\pi$ to $\pi$ and meta-atom size jumps from smallest to largest), as well as the metaoptic edges. However, the NN method provides excellent quantitative agreement with the full-wave prediction, especially in estimating the field intensity at the focal point. **Figure 2**c quantifies the irradiance at the focal plane as a function of f/# for each metaoptic design method. It illustrates the consistent accuracy of the NN across f/#'s, as well as the drop-off in accuracy of the LPA prediction method as the f/# decreases. This loss in accuracy can be explained twofold - (1) neighboring meta-atoms' radii change more rapidly in stronger optics (lower f/#) to the faster phase variation, meaning the meta-atom will behave less like the meta-atom in an infinite array, and (2) an increased number of Fresnel zone edges, at which the jump from largest to smallest meta atom causes a strong divergence from the meta-atom library behavior due to a larger field discontinuity. The noted average percent difference across all considered f/# lenses, relative to the full-wave (ground truth) method, for the ideal, LPA, and NN are 62.2%, 21.1% and 0.47%, respectively.

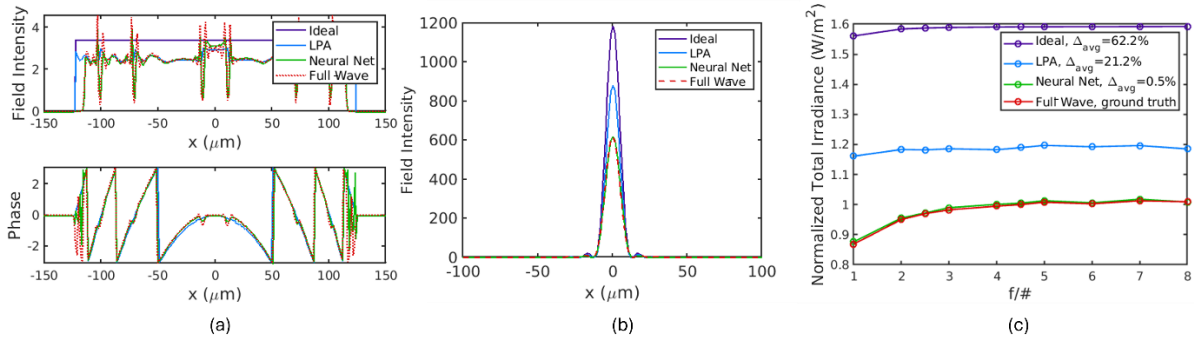

**Figure 2.** Validation of neural network method through comparison with ideal (Zemax), LPA, and full wave (Lumerical) methods; a) cross section of field amplitude and phase immediately after an f/2.5 metaoptic; b) field intensity cross-section at the focal point of an f/2.5 metaoptic; c) irradiance at focal planes for f/1 to f/8 metaoptics, normalized to the irradiance of full wave f/4.5; legend includes average percent difference compared to full wave.

Finally, to demonstrate the performance of the transformer-based NN method in a macroscale optical system with multiple powered optical surfaces, we constructed a singlet with two metaoptic surfaces in OpticStudio (**Figure 3**). The optical chain consisted of a silicon substrate with a metaoptic nanopatterned onto each surface of the substrate, as illustrated in **Figure 3**a; the metaoptic diameters are 0.5 mm and 0.244 mm respectively, remaining relatively small to computationally enable a comparison to ground truth (full-wave simulation). A separate NN was trained for the beam propagating air-metaoptic-silicon and silicon-metaoptic-air scenarios because the change in index results in a different field transformation as well as different available diffractive orders. In this system, the field incident on each metaoptic is not necessarily a flat plane wave. Because the NN was trained only on normal incidence plane wave illumination, we developed a "surrogate" method to calculate metaoptic performance that accounts for non-normal incidence, non-uniform illumination. This method can be applied separately to each metaoptic and consists of multiplying the incident field on the metaoptic by the same metaoptic's output field based on normal incidence illumination. We verified the accuracy of this method using off-normal and Gaussian illumination of metaoptics in Lumerical FDTD and these results can be found in the Supporting Information. This field multiplication method is a good approximation due to linear optics, although there is some margin for error due to the slight angular dependence of the meta-atoms (see Supporting Information).

To quantify the performance of each of the metaoptic simulation methods in our multi-metaoptic system, we compare the (1) optical performance via the irradiance onto the focal plane (**Figure 3**a), and (2) computation time (**Figure 3**b). Propagation between different optics in all cases was handled by OpticStudio's POP tool. In all cases, the source entering the optical system is a top hat. For the ideal case (purple), both metaoptics are treated as binary2 phase surfaces and all computation is completed within OpticStudio. In this ideal case, the calculation does not account for any optical loss within the metaoptics, leading to the highest calculated irradiance at the focal point. For all other methods, propagation through the metaoptic calculation is computed externally, by exporting the field 0.5 μm before the metaoptic in OpticStudio, applying one of the metaoptic methods, and then re-importing the calculated field 0.5 μm after the metaoptic; this integration of methods is performed automatically using OpticStudio's Python API. For the full-wave calculations, we considered two different scenarios, which we refer to as the full-wave import (red) and full-wave surrogate (yellow) methods. For the full-wave import method, the field exported from OpticStudio was imported directly into Lumerical as a custom import source, then the simulated field was exported back to OpticStudio for continued propagation. Consequently, this full-wave import method was treated as our ground truth scenario. As the name indicates, the full-wave surrogate method used the surrogate method described above; the full-wave simulation was run with a normal incidence plane wave and the resulting field was multiplied by the incident field from OpticStudio and then exported back into OpticStudio. We include this full-wave surrogate case to distinguish inaccuracies from the surrogate method versus the NN; note that this surrogate method is also used for the LPA prediction. As expected, the full-wave surrogate is quite close to the full-wave import (ground truth). For the case where the metaoptic output fields were generated using LPA (blue line), the accuracy is better than the ideal case, but a significant deviation is still noted relative to the full-wave import case, with the focal plane irradiance being overestimated by more than a factor of 2. However, in the case where the metaoptic output fields were predicted via the NN (green), only a minor deviation was observed, with predicted performance approaching that of both full-wave scenarios. Notably, simulation of both metaoptics in FDTD required on the order of a week using a compute cluster, whereas all other methods were completed within minutes on a workstation (**Figure 3**b), with most of the calculation time resulting from the handling and passing of data between OpticStudio and the chosen method. The calculation time of the NN itself is relatively quick, adding only an extra minute of compute time relative to the LPA method. This reduction in calculation time from weeks (full-wave) to minutes (NN) is a significant, many orders of magnitude improvement and helps circumvent the general intractability associated with simulating and calculating the performance of larger metaoptics accurately.

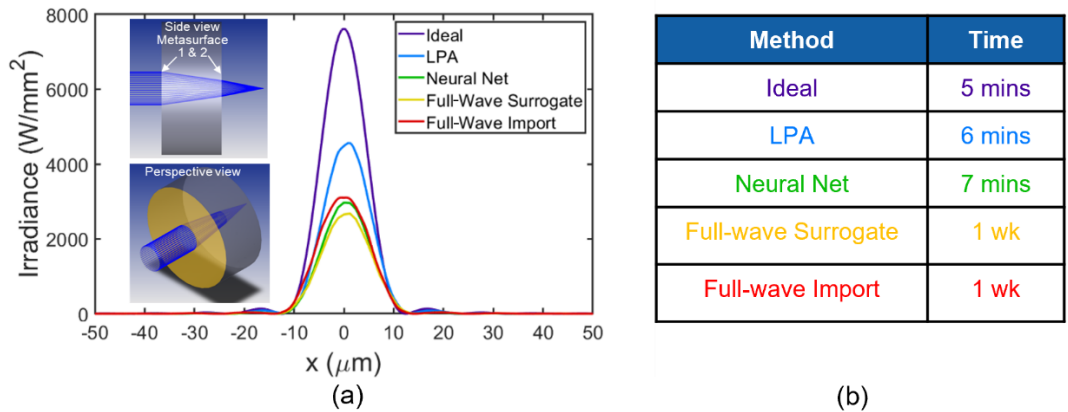

**Figure 3. (a)** Representative example optical system used to demonstrate performance of transformer-based neural network method, which consists of a silicon substrate with both sides nanopatterned with a metaoptic. **(b)** Comparison of optical performance is quantified by the irradiance onto the focal plane.

We proposed a new methodology for the forward design of optical systems containing metaoptics, using a surrogate transformer-based transformer NN solver integrated with a commercial ray optics software tool that performs beam propagation through Fourier optics propagation. We demonstrated its superior performance to traditional approximate methods with a two metaoptic optical system example, illustrating our method's utility as a solution to the tradeoff between speed versus accuracy. We validated the accuracy of the NN and compared the accuracy of our technique with the irradiance at the focal plane as our performance metric. The NN solver exhibited close quantitative performance to full-wave simulations with the speed of non-full-wave methods. Even for a system incorporating metaoptics that are <1 mm diameter in size at a wavelength of 4 μm, FDTD simulations require ~1 week on 100s of GB of RAM, while the NN predicts fields within minutes. Especially in the case where metaoptics need to be integrated into conventional optical systems, traditional FDTD techniques make simulation of metaoptic components computationally intractable, while the NN technique does not exhibit such a limitation. Consequently, this technique enables fast and accurate modeling of full optical systems containing metaoptics, which could enable optimization of metaoptics for improved performance and expedite their use in practical applications.